\documentclass[acmsmall,screen]{acmart}
\AtBeginDocument{%
  }

\setcopyright{acmlicensed}
\copyrightyear{2026}
\acmYear{2026}
\acmDOI{XXXXXXX.XXXXXXX}

\usepackage{tcolorbox}
\usepackage{todonotes}
\usepackage[T1]{fontenc}
\usepackage{xcolor}
\usepackage{tcolorbox}
\usepackage{listings}

\usepackage{zi4} %
\usepackage{upquote}

\usepackage{xspace}

\lstdefinestyle{pybox}{
  language=Python,
  basicstyle=\ttfamily\small,   %
  columns=fullflexible,
  keepspaces=true,
  numbers=left,
  numberstyle=\scriptsize\color{gray},
  stepnumber=1,
  numbersep=6pt,
  showstringspaces=false,
  breaklines=true,
  breakatwhitespace=true,
  tabsize=4,
  backgroundcolor=\color{gray!7},
  keywordstyle=\bfseries\color{blue!60!black},
  commentstyle=\itshape\color{red!55!black},
  stringstyle=\color{orange!70!black},
  frame=single,
  rulecolor=\color{gray!40},
  framerule=0.3pt,
  xleftmargin=1.4em,
  framexleftmargin=1.2em,
  aboveskip=6pt,
  belowskip=6pt,
}

\begin{document}

\title{\textsc{Echo}: Graph-Enhanced Retrieval and Execution Feedback for Issue Reproduction Test Generation}

\author{Zhiwei Fei}
\email{zhiweifei@smail.nju.edu.cn}
\affiliation{%
  \institution{Nanjing University}
  \city{Nanjing}
  \country{China}
}

\author{Yue Pan}
\email{jack.pan.23@ucl.ac.uk}
\author{Federica Sarro}
\email{f.sarro@ucl.ac.uk}
\affiliation{%
  \institution{University College London}
  \city{London}
  \country{United Kingdom}
}

\author{Jidong Ge}
\email{gjd@nju.edu.cn}
\affiliation{%
  \institution{Nanjing University}
  \city{Nanjing}
  \country{China}
}

\author{Marc Liu}
\email{marc@dplabs.io}
\affiliation{%
  \institution{NOFA}
  \city{Hong Kong}
  \country{China}
}

\author{Vincent Ng}
\email{vince@utdallas.edu}
\affiliation{%
  \institution{University of Texas at Dallas}
  \city{Richardson}
  \country{United States}
}

\author{He Ye}
\authornote{\textbf{Corresponding author.}}
\email{he.ye@ucl.ac.uk}
\affiliation{%
  \institution{University College London}
  \city{London}
  \country{United Kingdom}
}

\renewcommand{\shortauthors}{Zhiwei Fei et al.}

\begin{abstract}

Identifying the root cause of a bug remains difficult for many developers because bug reports often lack a bug reproducing test case that reliably triggers the failure. Manually writing such test cases is time-consuming and requires substantial effort to understand the codebase and isolate the failing behavior. To address this challenge, we propose \textsc{Echo}, an agent for generating issue reproducing test cases, which advances previous work in several ways. %
During generation, \textsc{Echo} strengthens context retrieval by leveraging a code graph and a novel automatic query-refinement strategy. \textsc{Echo} also improves upon previous tools by automatically executing generated test cases, a first-of-its-kind feature that seamlessly integrates into practical development workflows. In addition, \textsc{Echo} generates potential patches and uses the patched version to validate whether a candidate test meets the fail-to-pass criterion and to provide actionable feedback for refinement. Unlike prior bug-reproduction agents that sample and rank multiple candidate tests, \textsc{Echo} generates a single test per issue, offering a better cost--performance trade-off. Experiments on SWT-Bench Verified show that \textsc{Echo} establishes a new state of the art among open-source approaches, achieving a 66.28\% success rate.

\end{abstract}




\setcopyright{none}
\maketitle

\section{Introduction}
With the widespread adoption of GitHub, developers spend a lot of time dealing with issues reported by users. A critical step in issue resolution is \textit{issue reproduction}, which involves automatically generating executable code to reproduce the problem reported by the user in the issue description. Specifically, given an issue from a code repository that describes the problem and may include failed logs and steps to reproduce, the goal is to generate a test case that replicate the issue (from now on we will refer to this test case as the \textit{issue-reproducing test case}).
The generated issue-reproducing test case must satisfy the \textit{fail-to-pass criterion,} meaning that the test should fail in the original codebase and pass in the patched version. Successful issue reproduction not only accelerates problem localization and resolution, but also strengthens quality assurance in continuous integration and delivery, thereby improving the robustness and reliability of software systems.

Recent research~\cite {kang2023large,wang2025aegis,nashid2025issue2test,khatib2025assertflip,ahmedotter, ahmed2025execution} has begun automating the generation of issue-reproducing test cases leveraging Large Language Models (LLMs).
Existing proposals follow a similar pipeline, summarized as follows: (1) The approach first analyzes the issue to form hypotheses about the failure mode (e.g., through root-cause analysis) and to identify the files and functions likely involved~\cite{nashid2025issue2test}; (2) it then retrieves relevant context~\cite{nashid2025issue2test, ahmedotter, ahmed2025execution} from the repository, commonly including focal functions and regression tests, which provide useful information to understand the issue and the assertion styles needed to generate the issue-reproducing test case; (3) next, it synthesizes one or more candidate reproduction tests and inserts them into an appropriate test file (or creates a new test file) following the project’s testing framework; (4) finally, the candidates are executed and validated (often using runtime feedback~\cite{nashid2025issue2test} or dual-version checks~\cite{ahmed2025execution}) to confirm that the observed failures align with the issue report and that the test satisfies the fail-to-pass criterion. Some approaches iterate this loop for refinement, while others generate multiple candidates and select the best one using execution-based criteria. %

Within this pipeline, accurately retrieving relevant code and regression tests is essential for issue reproduction, but existing approaches have notable limitations. Most rely on the Agentless framework \cite{xia2025demystifying}, which uses a simple hierarchical LLM-based file localization method. While easy to integrate, it lacks the ability to capture deeper structural relationships and long-range semantics in large codebases, often resulting in noisy or incomplete context and reduced reproduction performance.
Additionally, many test generation approaches assume that test execution commands and configurations are already known. This assumption holds in benchmarks with standardized harnesses but breaks down in real-world repositories, where test execution details are implicit and scattered across multiple files. Even small command mismatches can prevent tests from running correctly, making execution setup a major open challenge.
Finally, issue reproduction suffers from the absence of reliable failure oracles. Although issue descriptions may include logs or stack traces, they rarely provide an executable way to verify that a generated test fails for the correct reason. As a result, systems often rely on LLMs to judge semantic equivalence between failures and issue descriptions, which is inherently unreliable.

To address these challenges, we propose \textsc{Echo}, a novel agent for reproduction test case generation that leverages a code graph to enhance context retrieval. Code-graph representations~\cite{liu2025codexgraph,liu2024graphcoder,chen2025prometheus, tao2025code, athale2025knowledge} have shown clear advantages over simple LLM-based project-level retrieval, especially when the goal is to precisely localize relevant files and tests. Leveraging the code graph, \textsc{Echo}’s context retrieval component automatically refines retrieval queries, iteratively checks whether the collected context is sufficient to generate the test, and finally returns a compact set of context. After generating a candidate test, \textsc{Echo} automatically synthesizes the appropriate test command for the target project, runs the test, and collects execution feedback. In the verification stage, \textsc{Echo} does not rely solely on LLM-based semantic judgment. Instead, building on the key idea of e-Otter++~\cite{ahmed2025execution}\footnote{e-Otter++~\cite{ahmed2025execution} is the first work leveraging execution feedback to select among generated candidates using the patched version as an oracle. It uses the potential patches produced by the Agentless tool, treats these patched versions as references, and then generates a pool of candidate reproduction tests. Finally, it selects the best test by checking whether it exhibits the desired fail-to-pass behavior across the buggy and patched versions. This strategy has shown promising results and state-of-the-art performance on SWT-Bench.}, \textsc{Echo} generates possible patch candidates and using the patched version as an oracle; that is, \textsc{Echo} checks whether the generated test satisfies the fail-to-pass criterion across the buggy and patched versions. \textsc{Echo} further enhances this process by using this information to refine the generated test case: If the test fails the fail-to-pass validation, \textsc{Echo} feeds the failing logs back to the generator and regenerates an improved test. While many prior approaches generate a large pool of candidates and then select the best one, \textsc{Echo} aims to be more efficient by focusing on producing a single high-quality reproduction test per issue through iterative refinement. 

Based on empirical results on SWT-Bench Verified (henceforth SWT-Bench-V), a subset of the widely-used SWT-Bench benchmark \cite{mundler2024swt}, we observe that \textsc{Echo} places itself as the new state-of-the-art (SOTA) among open-sourced bug reproduction tools. While most current work remains closed-sourced, 
we make \textsc{Echo} publicly available. %
Our contributions can be summarized as follows:

\noindent (1) We introduce \textsc{Echo}, a novel issue reproduction agent that leverages a code graph to enhance context retrieval, uses execution feedback for refinement, and can automatically execute the generated tests.

\noindent(2) We provide empirical evidence posing \textsc{Echo} as new SOTA, with a 66.28\% success rate on SWT-Bench-V. We further analyze the execution-based refinement component and cost, and summarize practical pitfalls that help explain when and why \textsc{Echo} succeeds or fails.

\noindent(3) We open-source \textsc{Echo} to facilitate its adoption and support reproducibility and future research. We make our code, execution logs, and full experimental results publicly available at \url{https://github.com/EuniAI/Echo}.

\section{Motivating Example}\label{sec:motivating}

\begin{figure}[b]
\centering
\small
\begin{tcolorbox}[
  sharp corners,
  colback=white,
  colframe=black,
  boxrule=0.6pt,
left=3pt,right=3pt,top=3pt,bottom=3pt
]
\textbf{Title}:
Cannot override get\_FOO\_display() in Django 2.2+\par\smallskip

\textbf{Description}:
I cannot override the get\_FIELD\_display function on models since version 2.2.
It works in version 2.1.\par\smallskip

\textbf{Example:}\par
\begin{lstlisting}[style=pybox,basicstyle=\footnotesize]
class FooBar(models.Model):
    foo_bar = models.CharField(_("foo"), choices=[(1, 'foo'), (2, 'bar')])

    def __str__(self):
        return self.get_foo_bar_display()  # This
        # returns 'foo' or 'bar' in 2.2, but 'something' in 2.1

    def get_foo_bar_display(self):
        return "something"
\end{lstlisting}

What I expect is that I should be able to override this function.
\end{tcolorbox}

\caption{Django-12284 issue from SWT-Bench.}
\label{fig:test_case_generation_prompt}
\end{figure}

In this section, we present a motivating example that illustrates the challenges of generating reproduction test cases. We consider a real issue from SWT-Bench-V (Django-12284), shown in Figure \ref{fig:test_case_generation_prompt}. In this issue, the developer reports that the get\_FIELD\_display function no longer works as expected and provides only a brief code snippet.

The first challenge is that issue reports are often ambiguous and underspecified, making it difficult to understand the root cause and to infer what information is necessary for constructing a correct reproduction. As shown in Figure \ref{fig:test_case_generation_prompt}, the code snippets in the issue do not include required imports or a failing test case, and it does not clearly state the expected behavior or the minimal conditions under which the failure occurs. Consequently, a model (or a developer) must guess which parts of the repository context are relevant (e.g., which functions, or input, triggers the behavior) and which are irrelevant. Traditional context-retrieval pipelines typically rely on human-designed heuristics (e.g., fixed query or manually curated localization rules) to select candidate files and functions, but such heuristics do not always align with what an LLM needs to generate an executable test case.

A second difficulty lies in translating the natural-language issue description (plus a partial snippet) into runnable code that integrates into the project’s existing test framework. The report describes the observed failure, but it does not provide a concrete test harness. To create a valid reproduction test, the generator must (1) choose the correct testing style used in the repository (e.g., test framework used in Django), (2) place the test in the right location with correct naming and imports, and (3) construct any required environment or minimal setup (such as activate the execution environment, or test settings) so the snippet becomes an executable test. Given only the issue description, the task of recovering these repository-specific settings is non-trivial.

A third challenge is test execution. Different repositories adopt different commands, runners, and configuration requirements for executing tests (e.g., pytest vs. unittest, custom wrappers, environment variables, or database backends). Issue descriptions rarely include this information, yet it is essential for validating whether a generated test actually reproduces the problem. Even when the test code is correct, failing to execute the test code can prevent bug reproduction entirely.

Finally, there is often no reliable oracle for correctness. Given an ambiguous report, it is hard to formalize the buggy behavior and to determine whether a generated test truly matches the reported behavior, rather than failing for an unrelated reason. In practice, a useful reproduction test should fail for the same underlying reason as described in the issue and should pass once the bug is fixed. However, when the issue text does not precisely specify expected outcomes, determining this semantic alignment is itself a challenging verification problem. 

These challenges motivate the need for reproduction systems that (1) retrieve context in a way that matches the needs of code generation, (2) produce repository-conformant tests, (3) reliably discover how to execute them, and (4) validate reproduction using more reliable criteria.

\section{Related Work}\label{sec:related_work}
While some methods exist for generating reproduction tests~\cite{soltani2018search, wang2024application, wang2024feedback, zhao2022recdroid}, they are not designed to handle issues. Thus, in this section, we focus on current work using LLMs for issue-reproducing test case generation based on issues, which is the most relevant to our work.

LIBRO \cite{kang2023large} is the first approach to leverage LLMs for general bug reproduction. It samples the model multiple times to generate candidate tests, post-processes them into executable scripts for the target program, and then curates and ranks the most promising candidates to reduce developer inspection effort. AEGIS \cite{wang2025aegis} proposes a two-agent framework, where a searcher retrieves relevant context and a reproducer generates and iteratively refines test scripts. Its key contribution is a finite-state-machine controller that drives structured feedback loops using syntax checks, execution results, and external verification. Issue2Test \cite{nashid2025issue2test} introduces a three-phase pipeline that meta-prompts project-specific test-writing guidelines, performs root-cause analysis, and generates test candidates. It then runs an execution-feedback loop with two LLM roles: one classifies test failures, and the other verifies whether the observed assertion failures match the original bug report. AssertFlip \cite{khatib2025assertflip} assumes that LLMs are stronger at producing passing tests, so it first generates a test that passes under buggy behavior and then inverts it into a bug-revealing (fail-to-pass) test. Otter \cite{ahmedotter} uses self-reflective action planning with three components: a localizer that queries the LLM to identify relevant files, test functions, and focal functions, an action planner that iteratively refines a validated plan, and a test generator that produces complete test functions using localized code and structural cues (e.g., signatures and imports). Otter++ \cite{ahmedotter} improves Otter by generating multiple candidates via heterogeneous prompts that vary how much localized context is included. e-Otter++ \cite{ahmed2025execution} further extends Otter++ by selecting candidates using execution feedback and incorporating Agentless-generated patches, using the patched version to guide final test selection.

\begin{figure}[t]\centering\includegraphics[width=0.9\linewidth]{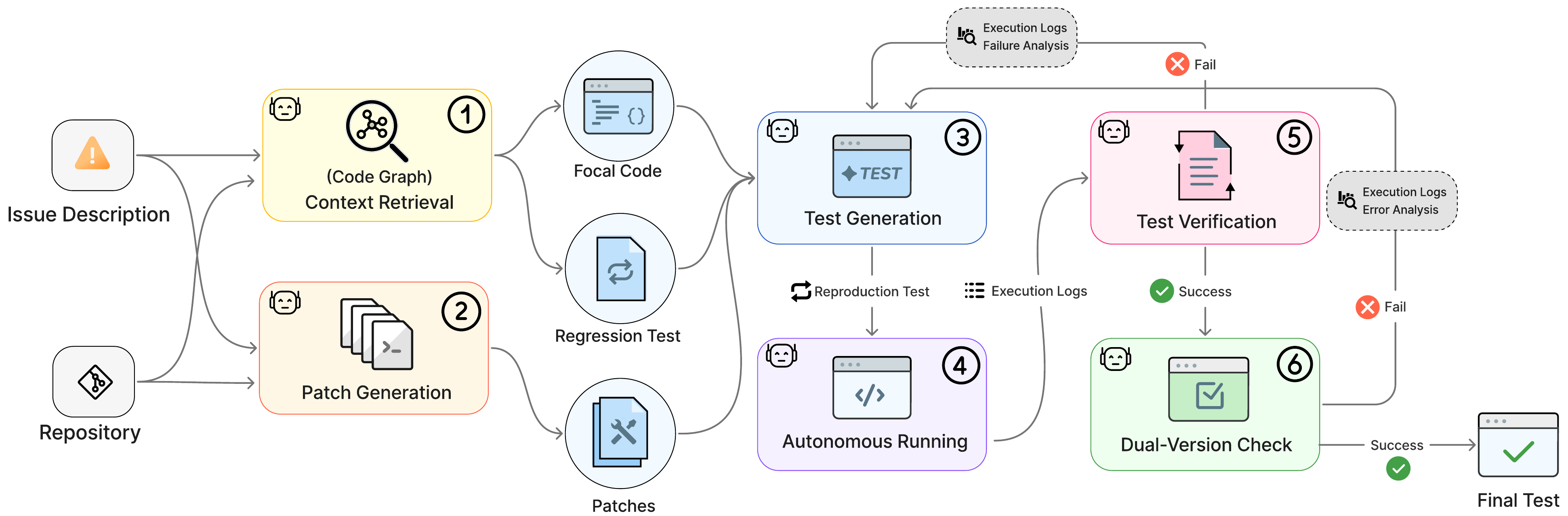}
    \caption{Overview of \textsc{Echo}.}
    \label{fig:Echo_overview}
\end{figure}

\section{Overview of \textsc{Echo}}\label{sec:approach}

\textsc{Echo} comprises six main phases: 1) context retrieval, 
2) patch generation,
3) test case generation, 
4) autonomous test case execution,
5) test verification, and
6) dual-version check.
Figure \ref{fig:Echo_overview} provides an overview of our method. \textsc{Echo} first converts a target repository into a code graph, uses LLMs to automatically refine the queries, and retrieves relevant code snippets and related regression tests to provide useful context for generating a reproduction test. Then \textsc{Echo} generates a potential patch to help generate a test case. After the LLM generates a candidate reproducing test, \textsc{Echo} uses a potential patch to construct a patched version of the project to verify the fail-to-pass criterion and, if necessary, refine the test based on execution feedback.
The details of \textsc{Echo} are described in the following subsections.

\subsection{Context Retrieval} 

Issues provided by developers are often ambiguous and usually lack detailed information about the failure. To mitigate this, context retrieval aims to search for useful information when generating the reproduction test case. In this phase, the code base is transformed into a code graph, and based on this code graph, we use queries to search for focal code and regression tests. Some ad hoc queries are not well designed, so we propose a self-improvement query-refinement mechanism that lets LLMs decide whether the query is good enough and the retrieved information sufficient for the task.

\subsubsection{Code Graph Construction}
We model the codebase as a heterogeneous directed graph with three node types: \texttt{ResourceNode}, \texttt{ParseNode}, and \texttt{TextSpanNode}.
A \texttt{ResourceNode} represents a file or directory, identified by a unique \texttt{node\_id}, and stores its \texttt{relative\_path} and \texttt{basename}.
A \texttt{ParseNode} represents a Tree-sitter parse-tree node in a source file, with a \texttt{node\_id}, \texttt{start\_line}/ \texttt{end\_line}, the covered code text (including comments), and a grammar \texttt{type}.
A \texttt{TextSpanNode} represents a chunk of extracted text (e.g., documentation or comments), with a \texttt{node\_id}, optional \texttt{meta\_data}, and the raw text.
We define five directed edge types to connect these nodes.
\textsc{HAS\_FILE} links a directory to its child files or subdirectories.
\textsc{HAS\_AST} links a source file to the root \texttt{ParseNode} of its parse tree, and \textsc{PARENT\_OF} captures parent--child relations within the tree.
\textsc{HAS\_TEXT} links a resource to its \texttt{TextSpanNode}s, and \textsc{NEXT\_CHUNK} connects adjacent text spans to preserve their order.
These nodes and edges jointly capture repository structure, syntax, and surrounding text for downstream tasks. The constructed graph is stored in Neo4j.

\subsubsection{Automatic Prompt Refinement}
The Context Retrieval component performs iterative, query-driven retrieval over a Neo4j-backed heterogeneous knowledge graph. %
In the retrieval phase, a (predefined) query is transformed into a contextual prompt and used to invoke structured tools implemented with Neo4j to locate relevant focal code and candidate regression tests via multiple lookup strategies (e.g., filename lookup, AST-pattern search, and documentation traversal). Because retrieval quality is highly sensitive to the query specification, a better-designed query typically improves the precision of the retrieved context. However, ad hoc, human-written queries do not generalize well across diverse user intents and may not be optimal for different LLMs. To mitigate this issue, we introduce an automatic query-refinement mechanism: The agent starts with an initial query, then in the selection phase uses an LLM to filter and rank candidate snippets for task relevance and token efficiency; if the selected context is deemed insufficient, the refinement phase triggers a follow-up query based on the LLM's assessment of the missing information, and repeats the cycle until sufficient evidence is collected. This dynamic process enables precise and focused context retrieval, thereby strengthening \textsc{Echo}'s reasoning about focal code and regression test.

\subsubsection{Focal Code Retrieval}

Based on the code graph and the context retrieval component, we first retrieve the \textit{focal code}, which is the most relevant to the issue. %
As highlighted by \citet{ahmedotter} looking at this code can help get a better understanding of the issue, and modifying this code can fix the bug.  \citet{ahmedotter} use the term focal function to refer to the functions exercised by the tests and likely locations for a fix. We extend this notion beyond functions to include the most relevant code (i.e., focal code).  We use the issue to retrieve the focal code.
To help models better understand the code, we ask the LLM to retrieve all necessary files and assemble a minimal but sufficient dependency closure around the focal code. Following the retrieval instructions, the model (1) identifies the key components mentioned in the issue, (2) includes their complete implementations and class definitions, (3) includes related code from the same module that affects the behavior and (4) follows imports to pull in dependent definitions that directly impact the bug. We explicitly ignore test files during this step to avoid noise, and we cap the retrieved context by prioritizing nodes with short graph distance to the anchors and high textual relevance, ensuring that the downstream generator receives concise yet actionable code context for fixing the issue.

\subsubsection{Regression Test Retrieval}
In the regression test retrieval phase, we prompt an LLM with the issue and ask it to locate three existing tests that are most relevant to the reported bug. The prompt explicitly guides the model to first analyze the issue characteristics, including the core functionality under test, key inputs and configurations, expected failure modes, and potential environmental dependencies. Based on this analysis, the model searches the repository’s test suite for candidates that test the same or closely related functionality.

To make the retrieved tests directly usable as high-quality examples, we require each returned candidate to be complete and self-contained. Concretely, the model must preserve the exact file path and line numbers, and include all necessary imports, the full test method body, any helper functions invoked by the test, and the complete setup for mocks and fixtures. This strict format ensures that the retrieved tests can be reliably used as in-context examples during test generation and refinement, without additional manual reconstruction. Finally, we enforce a priority order during retrieval: Tests covering the functionality described by the issue are preferred first, followed by those matching similar error conditions, then those exhibiting comparable mocking patterns, and finally those with similar assertion logic.

\subsection{Patch Generation}
\textsc{Echo} leverages candidate patches to guide the generation, selection, and improvement of the reproduction test case. The generated patch is represented as a standard diff file, which can be directly applied to the target codebase to construct a patched version. The patch generator itself is not tied to any specific repair technique. It can be produced by any automated program repair tool.

In our implementation, we adopt Prometheus~\cite{chen2025prometheus} as the patch generation component. We choose Prometheus for two reasons. First, it is open-source, allowing everyone to download and run it to reproduce our setup and generate candidate patches directly from issue information. Second, Prometheus has demonstrated strong performance on SWT-Bench-V: among open-source issue-resolution approaches, it ranks in the top five, suggesting that it provides a reliable backbone for generating patches for real-world issues. Accordingly, we use Prometheus as the patch generator throughout our experiments, generating a single candidate patch for each issue.

\subsection{Test Case Generation}
In this phase, the LLM generates a reproduction test case by leveraging the context retrieved in the previous step. Specifically, we construct a test-generation prompt that combines the issue description, the retrieved focal code, a small set of relevant regression tests, and the patch. Rather than asking the model to modify existing tests in-place, we require it to produce a separate, self-contained test file that can be added to the current test suite. This design avoids unintended changes to existing tests, reduces the risk of breaking current tests and unrelated behaviors, and makes the generated artifact easier to review and integrate. The generated test must include all necessary imports, minimal setup code, and a runnable test function or class consistent with the project’s conventions.
To improve generation quality, we adopt role prompting (role play)~\cite{shanahan2023role} to improve instruction-following and task performance in LLMs. We prompt the model to act as a Quality Assurance automation expert responsible for writing a single focused reproduction test. The prompt instructs the LLM to generate one minimal test case, using the fewest assertions necessary to demonstrate the reported failure. We further provide a set of requirements to constrain the output: the test must reuse the example from the issue when available, target only the core bug behavior (avoiding redundant coverage), and follow the style and structure of similar existing tests, including imports, fixtures, mocks, and assertions. We also include one illustrative example in the prompt to clarify the output format.

\subsection{Autonomous Test Execution}
After generating a candidate bug reproducing test, \textsc{Echo} must execute it to validate that it triggers the reported failure. We rely on an LLM to infer the appropriate execution command from the repository context (e.g., the README of the project, the project layout, the test framework, and the configuration files) and run the test inside a container.
However, giving the LLM unrestricted shell access is risky and can undermine evaluation fairness. In preliminary trials, we observed that an unconstrained agent may take shortcuts that invalidate the intended verification procedure, such as checking for or recreating the test file, modifying project files to “make the test pass/fail,” and running the entire test suite and reporting an unrelated failure. These behaviors confound the measurement of whether the generated test itself reproduces the issue, and they may also introduce destructive actions (e.g., deleting or overwriting files) that break reproducibility.

To ensure that execution will not affect the current files and %
the validation step, we enforce the following execution requirements in the prompt:

\noindent \textit{Do not check whether the test file exists.} The reproduction test file is guaranteed to exist at execution time; file existence checks are unnecessary and may lead the agent to regenerate/relocate the test.

\noindent \textit{Do not run the full test suite.} The executor must run only bug reproduction test file produced by \textsc{Echo}, rather than invoking repository-wide test commands that may fail for unrelated reasons.

\noindent \textit{Do not edit any files.} The executor is strictly read-only. It may inspect files in the repository to determine the correct command, but it must not modify source code, configuration files, or tests.

\noindent \textit{Assume all dependencies are installed.} We assume the environment has already been properly set up, so the executor should focus on selecting and running the correct test command and must not attempt any environment setup, package installation, or dependency repair.

\noindent \textit{Stop once the test executes.} As soon as the test runner successfully executes the reproduction test (regardless of whether it passes or fails), the executor terminates and reports the outcome.%

These constraints restrict the LLM to the intended task—identifying and running the correct command for the generated reproduction test, preventing behaviors that could modify the repository, take shortcuts, or produce unrelated execution feedback. The execution feedback will be given to the verified step to determine whether it is successfully reproduced or contains some errors.

\subsection{Test Case Verification}
After obtaining execution feedback, we use an LLM to judge whether the generated test has truly reproduced the reported bug or should be further refined. To help the model understand how to distinguish a correct reproduction from an incorrect one, we provide three fixed examples, including one successful case and two failed cases. The verifier is instructed to assess semantic alignment between the observed failure behavior in the execution logs and the issue description, rather than treating “the test fails” as sufficient evidence. If the verifier concludes that the execution outcome is consistent with the reported bug, we terminate the loop; otherwise, we feed the execution logs back to the test generator, which uses this feedback to revise the test in the next iteration.

\subsection{Dual-version Check}
The test case that passes test case verification is then analyzed through a dual-version check to determine the final result. Specifically, \textsc{Echo} first applies the patches collected in the patch generation phase to the codebase and uses the patched version to validate whether the generated test meets the fail-to-pass criterion. If the test satisfies this fail-to-pass criterion, \textsc{Echo} returns it as the final result. Otherwise, \textsc{Echo} feeds the execution logs back to the LLM and regenerates the test in the next round.
The dual-version check is entirely rule-based and does not involve any LLM. It first runs the candidate test on both the original codebase and the patched version. It then parses the logs, matching failures and successes using keywords such as \textit{pass} and \textit{ok} (and similar terms) to determine whether the test passes on the patched version.
If the test case fails the dual-version check, the execution logs will be fed back to the generator and it will be retried at most twice. If the test still does not meet the criterion after the two retries, \textsc{Echo} stops and returns the last generated test as the final result for a given issue.

\section{Experimental Design}\label{sec:methodology}
\subsection{Research Questions}
To assess the effectiveness of \textsc{Echo}, we execute it on a popular open-source benchmark, SWT-Bench%
\footnote{\url{https://swtbench.com/}} 
\cite{mundler2024swt}, %
and compare it with the top-performing approaches on the SWT-Bench leaderboard. We also assess specific advantages and pitfalls of \textsc{Echo} when compared to other approaches.
This constitutes our first research question: \textbf{RQ1:} How effective is \textsc{Echo} compared to previous work?

A core component of \textsc{Echo} is the use of the execution feedback to refine the generated test cases. Therefore we aim to have a detailed understanding of this component and study the extent to which it affects the final results. This motivates our second research question: \textbf{RQ2:} How effective is the execution feedback of \textsc{Echo}?

To investigate the cost of \textsc{Echo}, we analyze token consumption at each stage and the average price per instance, then compare it with other approaches. This analysis provides a detailed view of the cost–performance trade-offs, helping developers decide how to strike a balance between effectiveness and expense~\cite{williams2026reflection}. This motivates our third and last research question: \textbf{RQ3:} What are the token consumption and LLM invocation costs of \textsc{Echo}?

\subsection{Overview of Evaluation Methodology} %
To answer \textbf{RQ1}, we compare \textsc{Echo} against other agents publicly reported on SWT-Bench. 
This comparison is possible as SWT-Bench provides trajectories for these approaches, including which issues are successfully and unsuccessfully reproduced, as well as their test coverage. 
However, several top-ranked submissions on the leaderboard do not provide sufficient details. For instance, L*Agent v1 and OpenHands are listed on the benchmark, but we could not find a research article or website describing their methods or providing their trajectories. We therefore exclude these systems in our comparison and focus only on those methods with publicly available technical details or trajectories, namely OpenHands (GPT-5-mini),
e-Otter++ (Claude 3.7 Sonnet),
Amazon Q Developer Agent (v20250405-dev),
AssertFlip (GPT-4o),
Otter++ (GPT-4o), 
Otter (GPT-4o),
OpenHands (Claude 3.5 Sonnet),
LIBRO (GPT-4o) and
Zero-Shot Plus (GPT-4o + BM25).
Leveraging the information provided in SWT-Bench, we present Venn diagrams that summarize overlap and complementarity across agents, highlighting the instances each method uniquely solves as well as the shared failure cases.
Besides comparing our results with those of the top-performing approaches on the leaderboard, we also provide a detailed analysis of \textsc{Echo}'s results. 
We group the failures into different categories (e.g., assertion errors) to identify which phases of \textsc{Echo} they are most likely related to in practice—either the test does not correctly reproduce the bug, or the failure is caused by execution issues. This analysis helps clarify the strengths and limitations of each component, and we further present case studies to illustrate these failures.

To assess the effectiveness of execution feedback refinement (\textbf{RQ2}), we compare \textsc{Echo} against a variant that removes this component. Specifically, the variant takes the retrieved focal code, regression tests, and patch information as input, and then directly generates a candidate issue-reproducing test. Next, an autonomous test executor runs the generated test, and a verifier inspects the execution logs to judge whether the observed failure behavior semantically matches the issue report. Unlike \textsc{Echo}, this variant does not perform dual-version check and therefore does not use the execution information between buggy and patched versions to further refine the test case. We refer to the results produced by this variant as \textsc{Echo} w/o refinement.
We further conduct an analysis to quantify how many generated tests satisfy the fail-to-pass criterion on the first attempt, how many additional tests become valid after refinement, and how many candidates ultimately fail to meet the criterion. For each outcome category, we also report its relationship with test correctness. To this end, we parse and analyze the execution logs produced by the bug reproduction agent across all instances. Finally, we conduct a case study in which a test initially appears correct (i.e., its observed failure behavior aligns with the issue description under log-based verification) but fails the dual-version check, possibly because the reference patch is incorrect. We then illustrate how the agent handles this scenario and leverages dual-version feedback to refine the test.

To quantify the cost of \textsc{Echo} (\textbf{RQ3}), we parse the execution logs and report the number of input tokens, the number of output tokens, and the total number of tokens per instance. Token consumption is a key factor of LLM inference cost, yet many prior studies do not report it \cite{williams2026reflection}. For Gemini 2.5 Pro, we treat “thinking” tokens as output tokens and include them in the reported totals. Token-level reporting makes it possible to estimate costs across models using public pricing, and is more informative than reporting only the dollar cost per instance. In contrast, existing work typically reports only the overall cost or the average per-instance cost for a single backbone model, without a token-level breakdown. We also provide the cost per project and the cost per instance and compare them with 
those reported in the previous work. 

\subsection{Evaluation Dataset}
As mentioned above, we evaluate \textsc{Echo} on SWT-Bench \cite{mundler2024swt}, which is derived from SWE-Bench \cite{jimenezswe}, to assess whether a generated test reproduces the reported issue using the fail-to-pass criterion. 
SWT-Bench encompasses popular GitHub repositories, containing real-world issues, ground-truth bug-fixes, and golden tests \cite{mundler2024swt}. 
We choose SWT-Bench because it is widely adopted and provides a public leaderboard with released execution trajectories, which facilitates the analysis of the differences across models. SWT-Bench filters out instances whose golden patches are unreliable due to flaky tests and offers two subsets: SWT-Bench Lite (276 instances) and SWT-Bench Verified (433 instances). In particular, SWT-Bench Verified is used more frequently than SWT-Bench Lite, with more evaluated models and more available trajectories, which further supports our analysis. Therefore, we use SWT-Bench Verified as our evaluation dataset.

\subsection{Models}
We chose Gemini 2.5 Pro \cite{comanici2025gemini} as our backbone model because, at the time of our experiments (before September 10, 2025), it was among the strongest publicly available “thinking” models for code-intensive reasoning. Google reported that Gemini 2.5 Pro ranked Top\#1 on LMArena and achieved SOTA results across a wide range of reasoning and coding benchmarks, including 63.8\% on SWT-Bench-V under a custom agent setup. This capability is particularly important for issue reproduction, where the model must synthesize the issue description, patch context, and retrieved focal code and regression tests into an executable issue-reproducing test.
Gemini 2.5 Pro also offers an extremely large context window, enabling us to provide rich repository context in a single request. In the Gemini API, gemini-2.5-pro allows up to 1,048,576 input tokens and 65,536 output tokens, alongside features such as function calling, structured outputs, code execution, and context caching. This is essential for our task, which requires processing a large amount of context for analysis and producing a structured diff file of the issue-reproducing test case.
We set the temperature of Gemini 2.5 Pro to 0.5 and use the default values for all other parameters. %

\subsection{Evaluation Measures}
In our study, we use the evaluation measures proposed in SWT-Bench~\cite{mundler2024swt}, described as follows. Given an instance with the original repository state $R$, the golden bug-fix patch $X^{*}$, the original regression test suite $T_{R}$, and a generated test patch that introduces tests $T$:

\smallskip
\noindent
The \textbf{Success rate ($S$)} is the fraction of instances for which the generated tests reproduce the issue.
A generated test $t \in T$ is considered \emph{fail-to-pass} ($F\!\to\!P$) if it fails on the original codebase
but passes after applying the golden patch:
$t(R)=F$ and $t(R\circ X^{*})=P$.
A generated test set $T$ \emph{reproduces} the issue if it contains at least one $F\!\to\!P$ test and
no test fails after the fix is applied (i.e., no $\times\!\to\!F$ test).
In addition, SWT-Bench reports the fraction of instances for which at least one
$F\!\to\!P$, $F\!\to\!\times$, and $P\!\to\!P$ test is generated.

\smallskip
\noindent
The \textbf{Change coverage ($\Delta C$)} 
measures %
the increase in line coverage by the generated tests on the code changes introduced by the golden patch.
Let $C^{R}_{T}(l)\in \mathbb{Z}_{\ge 0}$ denote the number of times line $l$ is executed when running test suite $T$ on codebase $R$.
Let $X_r$ and $X_a$ denote the removed and added lines of a patch $X$, respectively.
To exclude non-executable changes (e.g., docs/config), define the \emph{executable} changed lines as: %
$X^{*}_{r} = \{\, l \in X_r \mid C^{R}_{T_R}(l) + C^{R}_{T^{*}}(l) > 0 \,\} \,and\ X^{*}_{a}= \{\, l \in X_a \mid C^{R\circ X}_{T_R}(l) + C^{R\circ X}_{T^{*}}(l) > 0 \,\},
$
where $T^{*}$ are the golden tests and (for SWT-Bench-V scoring) $X$ is instantiated as the golden patch $X^{*}$.
The \emph{change coverage} of generated tests $T$ w.r.t.\ patch $X$ is:
\begin{align}
\Delta C^{T}_{X} =
\frac{
\left|\left\{ l\in X^{*}_{r}\mid C^{R}_{T_R\cup T}(l) > C^{R}_{T_R}(l) \right\}\right|
+
\left|\left\{ l\in X^{*}_{a}\mid C^{R\circ X}_{T_R\cup T}(l) > C^{R\circ X}_{T_R}(l) \right\}\right|
}{
|X^{*}_{r}| + |X^{*}_{a}|
}.
\end{align}
Instances with $|X^{*}_{r}|+|X^{*}_{a}|=0$ are excluded from the coverage analysis.

\smallskip
\noindent
The \textbf{Patch well-formedness ($W$)} is the fraction of instances where the generated patch can be applied to $R$ without errors.
\section{Results}\label{sec:results}
\subsection{Answer to RQ1}
\subsubsection{\textsc{Echo}'s effectiveness} Table \ref{tab:Echo-results} summarizes \textsc{Echo}’s overall performance on SWT-Bench-V. \textsc{Echo} achieves an applicability of 93.76\%, indicating that most generated tests are well-formed and can be successfully integrated into the codebase. Most of the generated tests are potentially reproducing initially failing tests, as reflected by the 77.37\% F→X rate. %
\textsc{Echo} further achieves a success rate of 66.28\%, while maintaining a very low P→P rate of only 3.93\%, which implies that most generated test cases are closely related to the reported issues. Among the successful tests, the coverage delta reaches 76.50\%, indicating that these test cases cover most patched code and are thus of high quality.

\begin{table}[t]
  \centering
  \begin{minipage}[t]{0.48\linewidth}
    \centering
    \caption{RQ1. \textsc{Echo} results on SWT-Bench-V.}
    \label{tab:Echo-results}
    \begin{tabular}{l r}
      \hline
      Measure & \textsc{Echo} \\
      \hline
      Applicability (W) & 93.76 \\
      Success Rate (S) & 66.28 \\
      F$\rightarrow$X & 77.37 \\
      F$\rightarrow$P & 66.51 \\
      P$\rightarrow$P & 3.93 \\
      Coverage Delta ($\Delta^{all}$) & 58.54 \\
      Coverage Delta Resolved ($\Delta^{S}$) & 76.50 \\
      Coverage Delta Unresolved ($\Delta^{not\ S}$) & 22.50 \\
      \hline
    \end{tabular}
  \end{minipage}
  \hfill
  \begin{minipage}[t]{0.48\linewidth}
    \centering
    \caption{RQ2. \textsc{Echo} w/o refinement results.}
    \label{tab:Echo-first-metrics}
    \begin{tabular}{l r}
      \hline
      Measure & Results \\
      \hline
      Applicability (W) & 94.46 \\
      Success Rate (S) & 59.12 \\
      F$\rightarrow$X & 72.75 \\
      F$\rightarrow$P & 59.35 \\
      P$\rightarrow$P & 3.70 \\
      Coverage Delta ($\Delta^{all}$) & 53.83 \\
      Coverage Delta Resolved ($\Delta^{S}$) & 75.93 \\
      Coverage Delta Unresolved ($\Delta^{not\ S}$) & 23.12 \\
      \hline
    \end{tabular}
  \end{minipage}
\end{table}

Table \ref{tab:Echo_results_per_project_swt_verified} presents the effectiveness of \textsc{Echo} in generating F$\rightarrow$P test cases per project. The results are reported separately for the issues for which \textsc{Echo} successfully generated issue-reproducing test cases and for all issues present in SWT-Bench-V. \textsc{Echo} successfully generated a total of 419 issue-reproducing test cases; some instances were not generated because the reproduction process hit the step limit, as we require the LLM to produce an issue-reproducing test case within 500 steps for efficiency. Among the generated test cases, \textsc{Echo} achieves a 68.5\% success rate, successfully generating F$\rightarrow$P tests for 287 out of 419 issues, which is higher than the overall success rate on the full benchmark. Most hard issues that require more steps are in the {\tt django} project, with one in the {\tt pytest-dev} project. The results vary across projects. \textsc{Echo} performs best on scikit-learn, achieving a 91.67\% success rate, although this project has relatively few issues. Pydata also shows strong performance, with a 80.0\% F$\rightarrow$P rate. \textsc{Echo} achieves the worst performance on {\tt pallets} where it cannot generate any F$\rightarrow$P test case, followed by {\tt pylint-dev}, where it achieves only a 16.67\% F$\rightarrow$P rate, suggesting that it is challenging to generate effective test cases for these projects.

\begin{table}[tb]
\centering

\begin{minipage}[t]{0.49\textwidth}
  \centering
  \captionof{table}{RQ 1. \textsc{Echo} results per project on SWT-Bench-V.}\label{tab:Echo_results_per_project_swt_verified}
  \small
  \setlength{\tabcolsep}{4pt}
  \renewcommand{\arraystretch}{1.08}
  \resizebox{\linewidth}{!}{%
  \begin{tabular}{l|rrr|rrr}
    \toprule
    \textbf{Project} & \multicolumn{3}{c|}{\textbf{Tested issues}} & \multicolumn{3}{c}{\textbf{All issues}} \\
    \cmidrule(lr){2-4}\cmidrule(lr){5-7}
    & \textbf{Issues} & \textbf{F$\rightarrow$P} & \textbf{Acc. (\%)} &
      \textbf{Issues} & \textbf{F$\rightarrow$P} & \textbf{Acc. (\%)} \\
    \midrule
    django        & 203 & 133 & 65.52 & 216 & 133 & 61.57 \\
    sympy         & 73  & 57  & 78.08 & 73  & 57  & 78.08 \\
    scikit-learn  & 24  & 22  & 91.67 & 24  & 22  & 91.67 \\
    matplotlib    & 32  & 25  & 78.12 & 32  & 25  & 78.12 \\
    astropy       & 17  & 12  & 70.59 & 17  & 12  & 70.59 \\
    sphinx-doc    & 28  & 13  & 46.43 & 28  & 13  & 46.43 \\
    pydata        & 15  & 12  & 80.00 & 15  & 12  & 80.00 \\
    pytest-dev    & 14  & 10  & 71.43 & 15  & 10  & 66.67 \\
    pylint-dev    & 6   & 1   & 16.67 & 6   & 1   & 16.67 \\
    pallets       & 1   & 0   & 0.00  & 1   & 0   & 0.00  \\
    psf           & 4   & 1   & 25.00 & 4   & 1   & 25.00 \\
    mwaskom       & 2   & 1   & 50.00 & 2   & 1   & 50.00 \\
    \midrule
    \textbf{Total/Avg} & 419 & 287 & 68.50 & 433 & 287 & 66.28 \\
    \bottomrule
  \end{tabular}
    }
\end{minipage}
\hfill
\begin{minipage}[t]{0.5\textwidth}
  \centering
  \captionof{table}{RQ1. SWT-Bench-V official leaderboard results (success rate $\mathcal{S}$ and coverage increase $\Delta\mathcal{C}$).}
  \label{tab:swtbench-verified}
  \setlength{\tabcolsep}{4pt}
  \renewcommand{\arraystretch}{1.08}
  \resizebox{\linewidth}{!}{%
    \begin{tabular}{lrr}
      \toprule
      Approach & $\mathcal{S}$ (\%) & $\Delta\mathcal{C}$ (\%) \\
      \midrule
      \textsc{Echo} (Gemini 2.5 pro)                    & \textbf{66.3} & 58.5 \\
      OpenHands (GPT-5-mini)                           & 62.4 & 60.6 \\
      e-Otter++ (Claude 3.7 Sonnet)                    & 62.1 & \textbf{62.3} \\
      Amazon Q Developer Agent (v20250405-dev)         & 51.0 & 57.4 \\
      AssertFlip (GPT-4o)                              & 45.5 & 47.4 \\
      Otter++ (GPT-4o)                                 & 37.4 & 42.8 \\
      Otter (GPT-4o)                                   & 31.6 & 37.6 \\
      OpenHands (Claude 3.5 Sonnet)                    & 27.7 & 52.9 \\
      LIBRO (GPT-4o)                                   & 17.8 & 38.0 \\
      Zero-Shot Plus (GPT-4o + BM25)                   & 14.3 & 34.0 \\
      \bottomrule
    \end{tabular}
  }
  
\end{minipage}

\end{table}

\subsubsection{Comparison with Prior Work} 
\begin{figure}[t]
  \centering
  \begin{minipage}[t]{0.49\linewidth}
    \centering
\includegraphics[width=\linewidth]{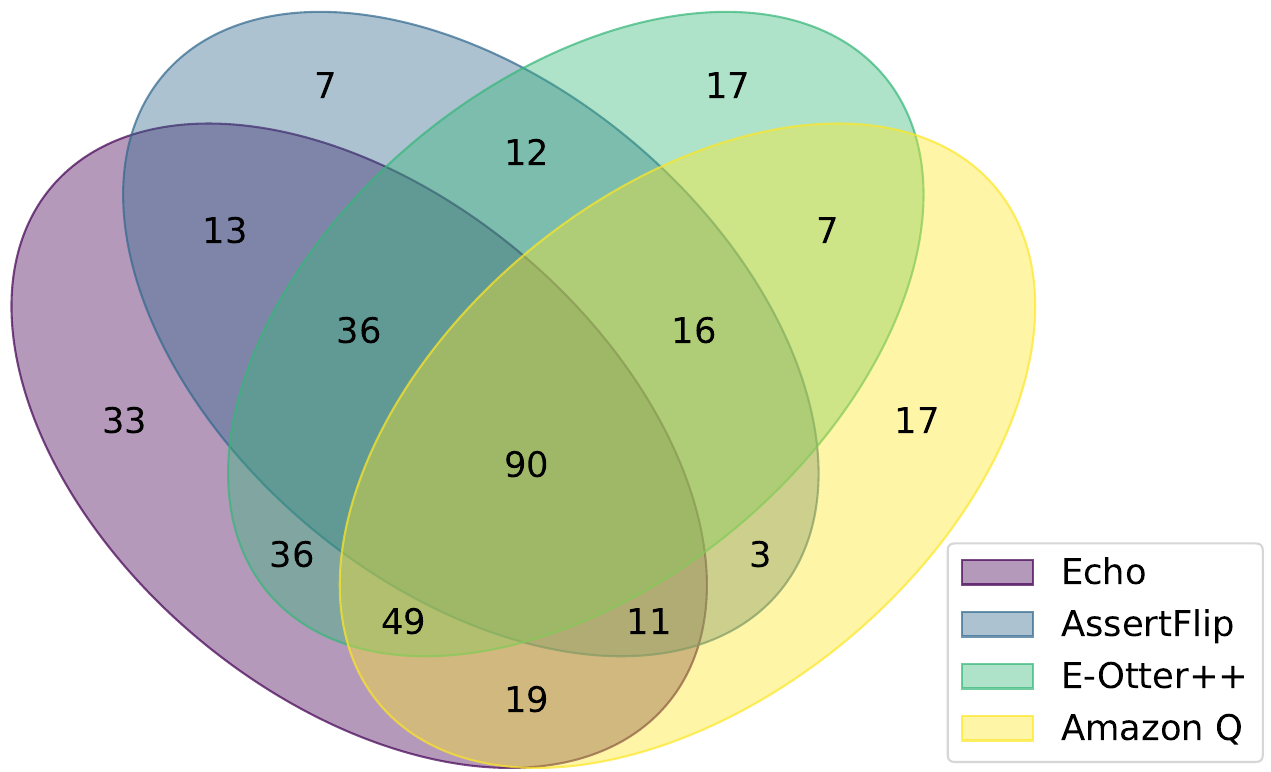}
    \caption{RQ1. Issues solved by the four approaches.}
\label{fig:venn_figure_of_four_approach}
  \end{minipage}\hfill
  \begin{minipage}[t]{0.49\linewidth}
    \centering
\includegraphics[width=\linewidth]{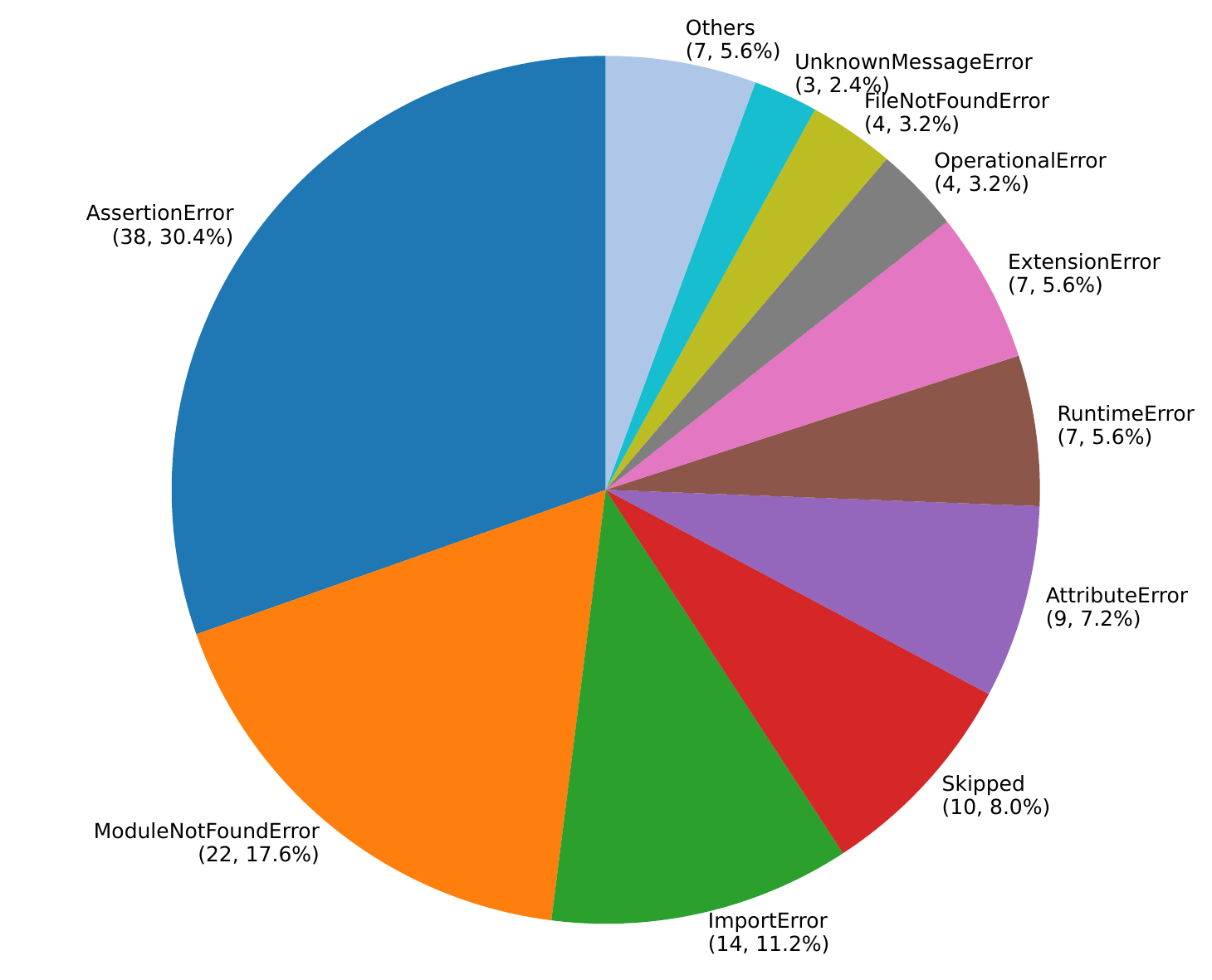}
    \caption{RQ1. Error distribution of unresolved instances.}\label{fig:unresolved_errors_pie_chart}
  \end{minipage}
\end{figure}

Table \ref{tab:swtbench-verified} compares the results of \textsc{Echo} with those of prior approaches as reported on the SWT-Bench-V leaderboard. We observe that \textsc{Echo} outperforms all other approaches on SWT-Bench-V: It surpasses OpenHands and e-Otter++, both of which use more advanced models (GPT-5-mini and Claude 3.7 Sonnet), by 3.9\% and 4.2\%, respectively. The improvements over older models are even greater, ranging from 15.3\% (vs. Amazon Q Developer Agent) to 52\% (vs. Zero-shot Plus).
Figure~\ref{fig:venn_figure_of_four_approach} shows the successfully solved instances for which issue-reproducing tests are generated by \textsc{Echo} and by three other approaches that achieve top performance and have published their trajectories on the leaderboard, namely AssertFlip, E-Otter++, and Amazon Q Developer Agent.
The diagram shows both the unique and overlapping sets of resolved instances. \textsc{Echo} achieves the largest number of unique tests, with 33 uniquely resolved instances that are not generated by any other approach. %
In contrast, AssertFlip produces only 7 unique tests, while E-Otter++ and Amazon Q each produce 17 unique tests. Overall, these results demonstrate that \textsc{Echo} is not only the best-performing approach for generating issue-reproducing tests, but also complements prior techniques by solving many instances that previous work cannot.

\subsubsection{Example of successful cases} Listing \ref{fig:Echo_sklearn_14141} provides an example test case that can only be generated by \textsc{Echo}, namely a test for the SWT-Bench-V instance \texttt{sphinx-doc\_\_sphinx-9281}. This issue concerns the readability of function signatures generated by Sphinx \texttt{autodoc} when a parameter has an \texttt{Enum} default value. Instead of displaying a concise reference such as \texttt{MyEnum.ValueA}, Sphinx renders the default using Python's verbose \texttt{repr} form, e.g., \texttt{<MyEnum.ValueA: 10>}, which clutters the documentation and is inconsistent with typical signature conventions.
\textsc{Echo} reproduces this issue with a minimal test that avoids running the full documentation build and instead targets the formatting step that \texttt{autodoc} relies on. It generated a test file with all necessary imports and a minimal test function. Specifically, the generated test defines a small \texttt{Enum} and checks the string produced for an enum member by Sphinx's inspection utility. The key assertion is that the description of \texttt{MyEnum.ValueA} should be the clean identifier \texttt{MyEnum.ValueA} rather than the verbose representation. On the buggy version, this check fails because the formatter returns the \texttt{repr}-style output, which later propagates into rendered signatures; after applying the fix, the formatter returns the concise name and the test passes. This confirms that the patch fixes the bug. %

\begin{lstlisting}[
  style=pybox,
basicstyle=\footnotesize,
  float=t,
  caption={Test case generated exclusively by Echo for \texttt{sphinx-doc\_\_sphinx-9281}.},
  label={fig:Echo_sklearn_14141}
]
import enum
import pytest
from sphinx.util import inspect

def test_enum_description():
    """Test the object_description() for enum values."""
    class MyEnum(enum.Enum):
        ValueA = 10
        ValueB = 20

    description = inspect.object_description(MyEnum.ValueA)
    assert description == "MyEnum.ValueA"
\end{lstlisting}

\subsubsection{Error Analysis}
We analyze the errors in cases where the generated test does not pass on the patched version under the SWT-Bench-V evaluation. We first conduct a detailed analysis of the \textit{Error Type}, which helps us understand how these tests fail and categorize the failure modes. We find that the errors can be grouped into 11 categories %
whose distribution is shown in Figure \ref{fig:unresolved_errors_pie_chart}. We observe that {\tt AssertionError} is the most common failure type, followed by {\tt ModuleNotFoundError}. {\tt ImportError} ranks third, and {\tt Skipped}, which indicates that the test case was not executed, is the fourth most frequent outcome. {\tt AttributeError} is the fifth most common failure type among these failed test cases. Together, these top five error categories account for 74.4\% of all failures.

A closer look reveals that most {\tt AssertionError}s occur because \textsc{Echo} does not correctly reproduce the reported behavior, and the generated tests fail on the patched version. {\tt ModuleNotFoundError} typically arises when the generated test imports modules that are not available in the testbed. For example, some testbeds do not include pytest, yet \textsc{Echo} may generate tests that import it at the beginning of the file. By examining the generation process, we find that the test generator does not precisely know the project’s runtime environment and therefore produces tests based on prior experience. In many cases, these tests fail because the execution process cannot run them successfully. In a few instances, the execution process attempts to modify the environment using pip install, which can also lead to {\tt ModuleNotFoundError}. Although we instruct the model not to install additional packages, it occasionally ignores this constraint, likely because the execution dialogue becomes long and the model loses track of earlier instructions. {\tt ImportError} is caused when a generated test attempts a relative import with no known parent package. For example, in \texttt{django\_\_django-14122}, the generated test uses \texttt{from .models import Author, Book, HardbackBook, Publisher, Store} to import the test objects. However, this relative import is not permitted in the SWT-Bench-V testing environment, and thus it triggers an {\tt ImportError}.

\begin{tcolorbox}[title = {Summary of RQ1}]
(1) \textsc{Echo} establishes itself as a new SOTA on SWT-Bench-V, outperforming all prior approaches, including methods with detailed technical reports and those without.

(2) \textsc{Echo} generates many unique issue-reproducing tests compared to existing approaches, suggesting that it can complement current methodologies.

(3) Most failures occur because \textsc{Echo} does not correctly generate issue-reproducing tests. In addition, mismatches between the execution environment and the project’s test environment can cause generated tests to fail. Test cases that use relative imports may also lead to failures.

\end{tcolorbox}

\subsection{Answer to RQ2}
We compare \textsc{Echo} with a variant that removes refinement, and we further break down how generated tests behave under the dual-version check. This allows us to quantify how often refinement is triggered, how many candidates are revised, and how this affects the final success rate. Dual-version feedback is not always reliable because the collected patch may be incorrect. In such cases, blindly following the execution feedback may sway the generator away from a correct test. What matters is whether the model can retain a test it believes is valid while discounting misleading feedback when necessary. We study this by identifying issues where \textsc{Echo} ultimately produces a successful regression test even though all intermediate candidates fail the dual-version check, and we present a case study that illustrates this situation in detail.

\subsubsection{Effect of Refinement}

\begin{figure}[t]
  \centering
  \begin{minipage}[t]{0.49\linewidth}
    \centering\includegraphics[width=0.70\linewidth]{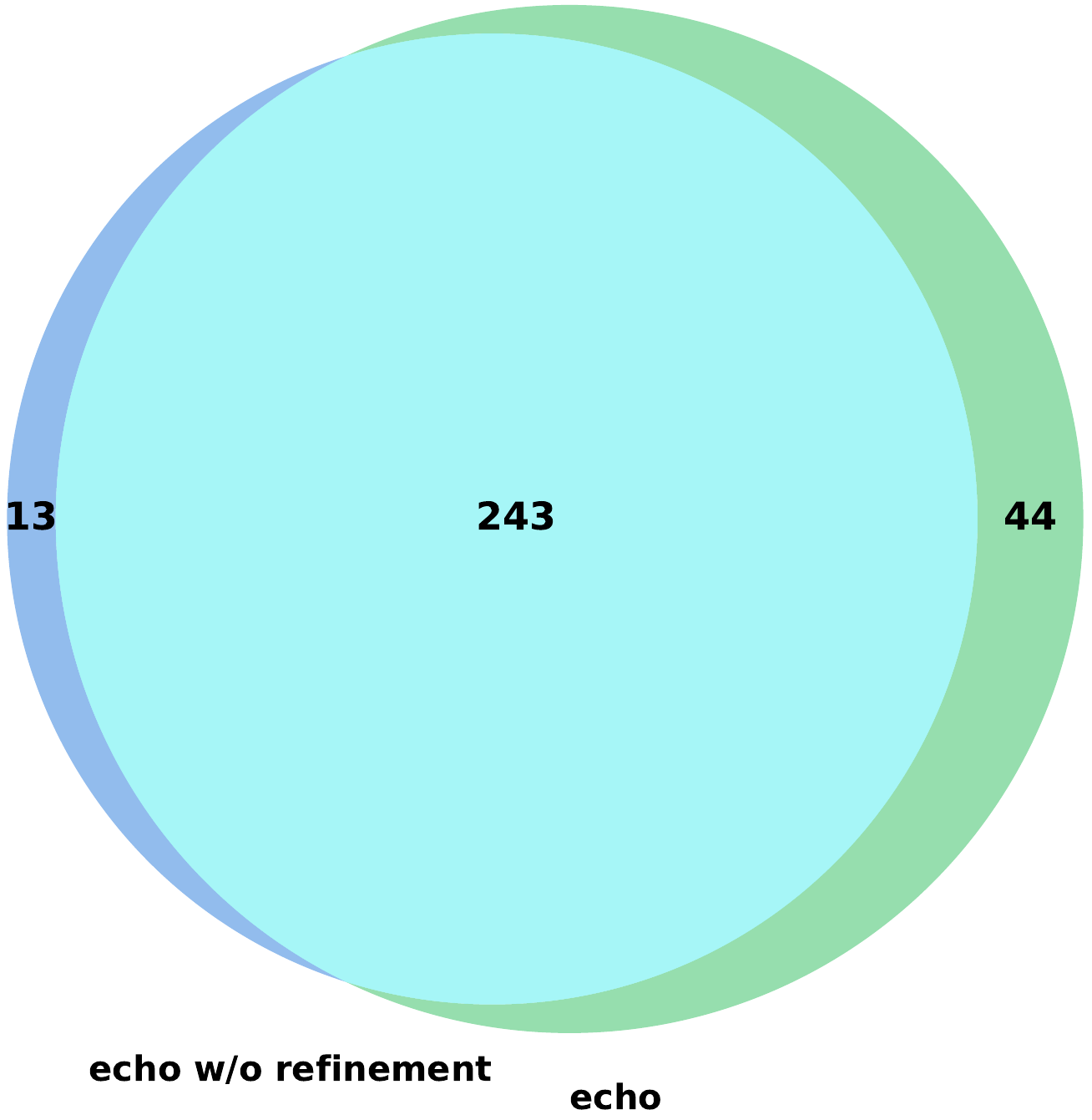}
    \caption{RQ2. Results of \textsc{Echo} with and w/o refinement.}
    \label{fig:venn_of_Echo}
  \end{minipage}\hfill
  \begin{minipage}[t]{0.49\linewidth}
    \centering\includegraphics[width=\linewidth]{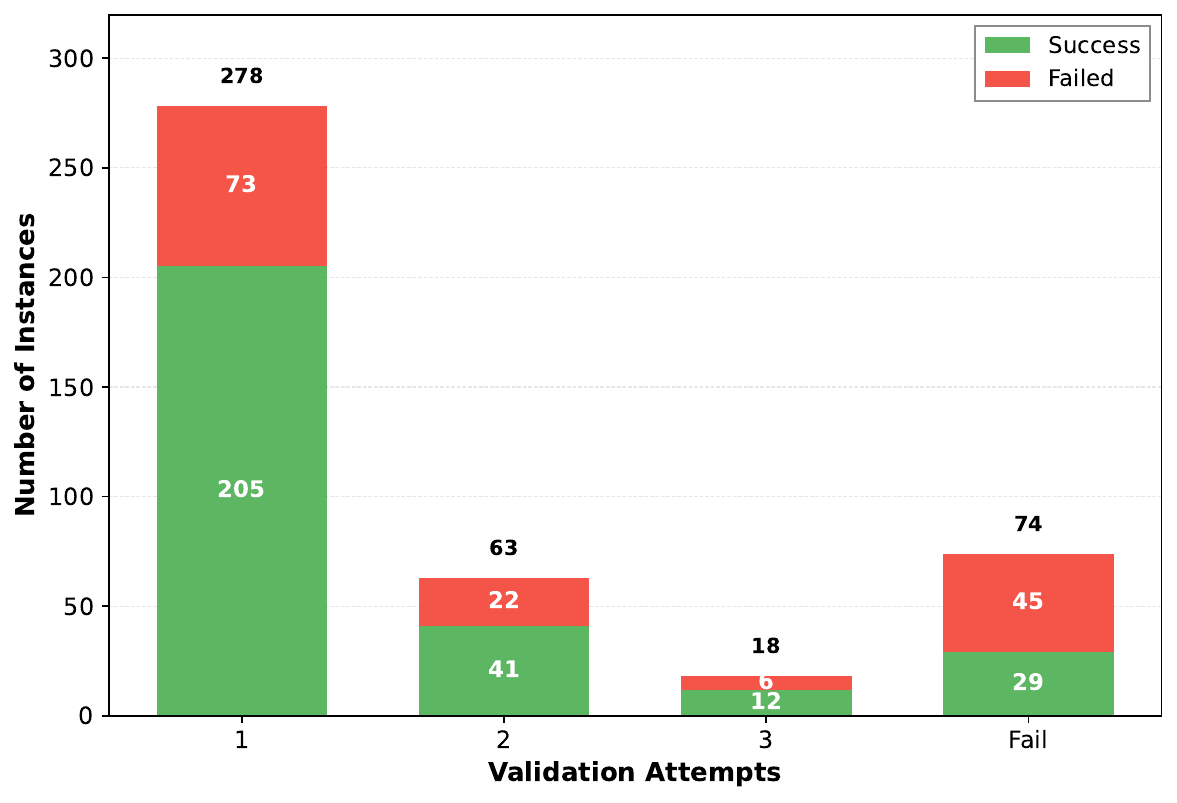}
    \caption{RQ2. Success rate of the dual-version check.}
    \label{fig:pass_rate}
  \end{minipage}
\end{figure}

Table \ref{tab:Echo-first-metrics} reports the results when we disable both the dual-version check and the refinement that leverages execution logs. In this setting, \textsc{Echo} reaches a 59.12\% success rate. Comparing these numbers with those in Table \ref{tab:Echo-results}, we observe a clear benefit from enabling dual-version checking and iterative refinement. The success rate increases by 7.16 percentage points, rising to 66.28\%. This suggests that validating candidates on the patched program and using execution feedback to refine the test helps the model produce higher-quality issue-reproducing test cases, and that the execution logs provide useful information for guiding the generation process.

We present a Venn diagram that illustrates how instances are solved across the two stages. As shown in Figure \ref{fig:venn_of_Echo}, the refinement process preserves most of the previously resolved instances: 243 remain successful after refinement. 44 additional instances succeed after refinement, while only 13 instances that were successful before refinement fail after refinement. Overall, these results indicate that the dual-version check and refinement are effective in improving performance.  

\subsubsection{Pass Rate Analysis} We perform a detailed analysis of the pass rate of the generated test case in the dual-version check. In this stage, \textsc{Echo} tries to verify the generated test case on the patched program and checks whether it passes on the this version. As explained in Section \ref{sec:approach} \textsc{Echo} makes two retry attempts. %
After that, if it still cannot meet the criteria, it outputs the last generated test case as the final version.  

Overall, 85.68\% of the generated test cases (359 out of 419) pass the dual-version check. Figure \ref{fig:pass_rate} shows the distribution of when verification is achieved. On the x-axis, '1' indicates that the first generated issue-reproducing test passes verification without any refinement, '2' indicates it passes after one round of regeneration, and '3' indicates it passes after two rounds. Failed means the test never passes the dual-version check. In this figure, "Success" refers to the tests that are true reproductions within the generated test case pool. 

A total of 278 generated test cases pass the dual-version check on the first attempt, and a large fraction of them are true issue-reproducing test cases (205). This suggests that, with graph-based retrieval and automatic prompt refinement, \textsc{Echo} can identify the most useful context and generate high-quality tests from it. For the test cases that do not initially pass the dual-version check, multi-round refinement enables some to meet the criterion, contributing many additional true reproductions.
Interestingly, even if a generated test case fails the dual-version check, it can still be correct, as shown in the last bar in Figure \ref{fig:pass_rate}. A closer inspection reveals two main reasons. First, some tests reproduce the issue, but the pipeline hits the step limit before reaching the dual-version check. This affects between 15 and 74 instances depending on the project. Second, among the remaining cases, although the refinement feedback may include incorrect execution logs, the LLM can distinguish this noise and preserve the correct issue-reproducing test it originally produced.

\subsubsection{Case Study} We provide a detailed example of the situation described above to better understand how LLMs behave in this setting. In many cases, the issue is caused by the execution environment rather than the generated test itself. After applying the patch, the working directory may change, and the LLM does not know it should switch to the correct test directory, which makes the execution command invalid. For example, in \texttt{django\_\_django-13741}, running the patched version can fail with an error such as \texttt{python: can't open file 'tests/runtests.py': [Errno 2] No such file or directory}.
When this happens, the LLM often focuses on the failure logs observed on the original codebase. If it confirms that the test already reproduces the issue, it tends to make only minimal edits to the test case, aiming to keep the test logic unchanged while adjusting formatting or structure details. This behavior helps preserve the correct issue-reproducing test even when the dual-version check is disrupted by environment related changes.

\begin{tcolorbox}[title = {Summary of RQ2}]

(1) Using execution feedback to refine the generated test cases is effective and improves the final results.

(2) LLMs can distinguish some misleading feedback and make only minimal changes, which helps preserve correct issue-reproducing tests.
\end{tcolorbox}

\subsection{Answer to RQ3}
We provide a detailed cost analysis of \textsc{Echo}. We first report the overall token usage and cost. \textsc{Echo}’s reproduction procedure can be divided into two stages: (1) information retrieval, which mainly collects the \emph{focal code} and \emph{relevant regression tests}, and (2) test generation and refinement, which uses the retrieved context to produce a candidate issue-reproducing test and iteratively refines it using execution feedback. We then break down the cost by stage to give a clearer picture of where tokens and money are spent, further separating the \emph{focal-code} retrieval and \emph{regression-test} retrieval components to understand their respective cost better. In addition, we analyze costs at the project level, as some projects may be more challenging for the model and therefore require more tokens and a higher expense. Finally, we compare \textsc{Echo} against publicly reported approaches to assess its cost-effectiveness.

\subsubsection{Overall Cost} We analyze token usage across 433 instances.
Overall, \textsc{Echo} consumes 400,251,303 tokens in total, with an average of 924,368 tokens per instance and a median of 470,113 tokens. This indicates that a small number of difficult instances in the dataset account for a vast majority of the cost. The distribution is heavy-tailed. The 95th percentile reaches 3,382,818 tokens.
The largest instance, django\_\_django-11734, consumes 21,433,984 tokens and is unsolved:
\textsc{Echo} tried its best to generate the reproduction test case and reach the maximum number of the steps and still failed to reproduce this issue.
\textsc{Echo}’s total cost across 433 instances is \$667.31, averaging \$1.54 per instance with a median cost of \$0.86.
The tail is heavy: the 95th percentile cost is \$5.20 and the 99th percentile cost is \$11.81, with rare outliers costing up to \$33.40. Overall, 80\% of the instances cost no more than \$1.95 and 90\% cost no more than \$3.01.

\subsubsection{Stage-level Cost} Table~\ref{tab:Echo-token-stage} summarizes the token usage and the cost per stage.
The  \textit{Gen \& Refine} stage plays a dominant role in the reproduction process, accounting for 69.7\% of all tokens and 71.6\% of the total cost. In contrast, focal-code retrieval and regression-test retrieval contribute 13.5\% and 16.8\% of the tokens respectively. In the information retrieval stage (code and test retrieval), the inputs account for 96.4\% of the stage’s token usage, while the outputs (including the thinking tokens) contribute only 3.6\%. This is because, in this stage, the model mainly judges whether the retrieved context is sufficient and thus produces little output for this decision and does not require detailed reasoning. 
In the \textit{Gen \& Refine} stage, the inputs account for 94.7\% of the tokens, while the outputs (including the thinking tokens) contribute the remaining 5.3\%. This is because \textit{Gen \& Refine} consumes all the context collected during retrieval and uses it to generate executable test code and iteratively refine it based on execution feedback. Concretely, the prompt includes focal code, regression tests, and patches, along with the intermediate artifacts produced during refinement (e.g., the candidate tests and the error logs), which makes the input extremely long. Compared to the retrieval stages, the number of thinking tokens increases sharply, as the model must perform deeper reasoning over the provided context and debug error logs to determine how to refine the issue-reproducing test case.

\begin{figure}[t]
  \centering
  \begin{minipage}[t]{0.49\linewidth}
    \centering
\includegraphics[width=\linewidth]{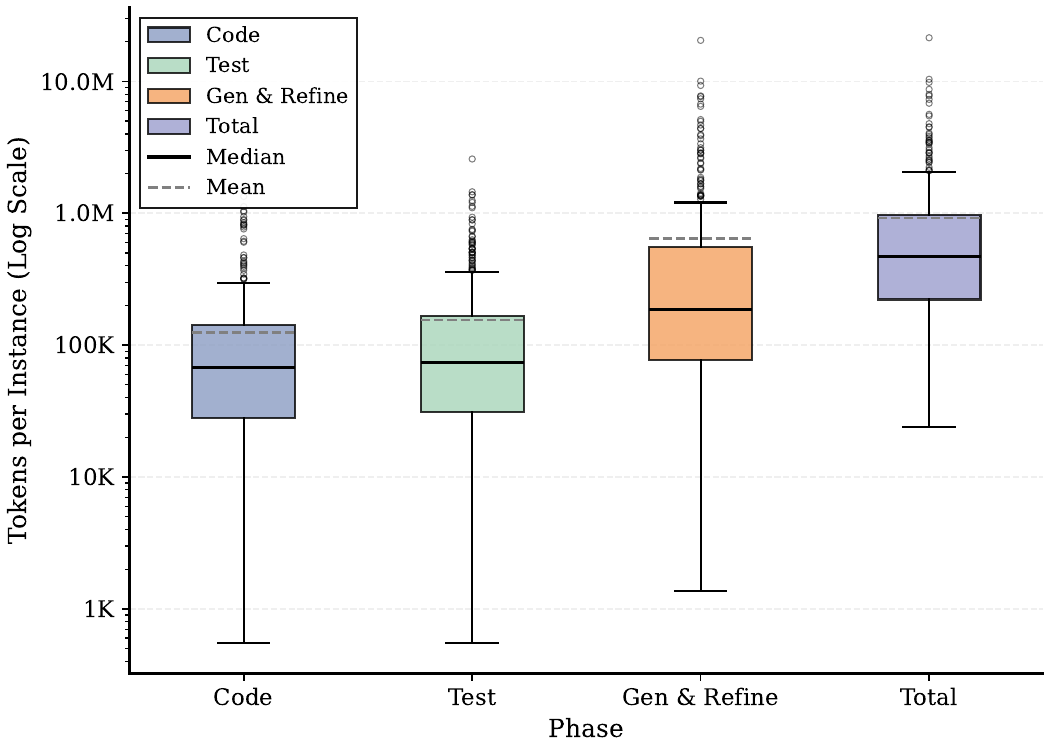}
    \caption{RQ3. Per-instance tokens per stage and overall. }
    \label{fig:Echo-token-dist}
  \end{minipage}\hfill
  \begin{minipage}[t]{0.49\linewidth}
    \centering
\includegraphics[width=\linewidth]{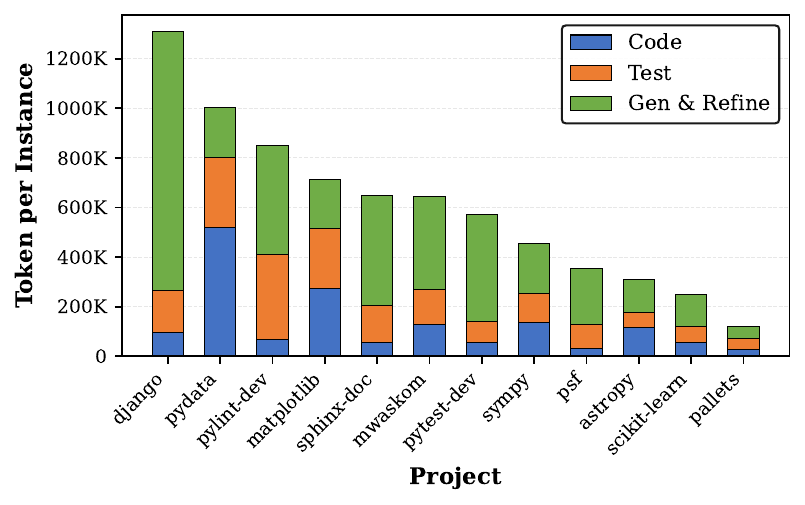}
    \caption{RQ3. Per-instance tokens per project.}
    \label{fig:Echo-project-mean}
  \end{minipage}
\end{figure}

Figure \ref{fig:Echo-token-dist} reports the per-instance token usage for each phase as well as the total on a log scale. The  \textit{Gen \& Refine} stage exhibits the largest variance and the most pronounced heavy tail, indicating substantial instance-to-instance difficulty differences: some issues are resolved with minimal generation effort and thus consume far fewer tokens and cost, whereas harder issues require much longer generation and refinement, dominating the high-token outliers.

\begin{table}[t]
  \centering
  \caption{RQ3. Token and cost summary per stage.}
  \label{tab:Echo-token-stage}
  \small
  \setlength{\tabcolsep}{4pt}
  \renewcommand{\arraystretch}{1.08}
  \begin{tabular}{lrrrrrr}
    \toprule
    Stage & Token share & Avg input & Avg output & Avg thinking & Avg total & Avg cost (\$) \\
    \midrule
    Code retrieval & 13.5\% & 120,297 & 911 & 3,677 & 124,885 & 0.20 \\
    Test retrieval & 16.8\% & 149,957 & 1,392 & 4,034 & 155,383 & 0.24 \\
    Gen \& Refine & 69.7\% & 610,038 & 8,140 & 25,922 & 644,100 & 1.10 \\
    \midrule
    Overall & 100.0\% & 880,291 & 10,443 & 33,633 & 924,368 & 1.54 \\
    \bottomrule
  \end{tabular}
\end{table}

\subsubsection{Project-level Cost}

Token consumption differs substantially across projects. Figure~\ref{fig:Echo-project-mean} shows the per-instance token usage by project. Django has the highest token usage per instance, followed by \texttt{pydata} and \texttt{pylint-dev}. The stage-level breakdown also varies across projects. For Django, \textsc{Echo} spends most tokens in the \textit{Gen \& Refine} stage, suggesting that many Django instances require more iterations and deeper analysis during test synthesis and debugging. In contrast, for \texttt{pydata}, code retrieval dominates token usage, which indicates that the issue descriptions are often underspecified and the agent must read more repository context to gather sufficient details for generating an issue-reproducing test. We further report the average cost per instance for each project. Consistent with the token analysis, Django is the most expensive project costing \$2.13 per instance: \textsc{Echo} sometimes struggles on the difficult instances and requires more steps, leading to a higher overall token usage and cost than in other projects. Pallets are the cheapest project costing only \$0.21 per instance.

\subsubsection{Comparison with other Methods}

We collect the costs of prior methods from the SWT-Bench paper~\cite{mundler2024swt}, which provides cost estimates for most baselines evaluated on the benchmark. On SWT-Bench Lite (276 instances) under GPT-4, the lowest-cost methods are the LLM-only baselines: \textsc{ZeroShotPlus} costs \$80.70 and \textsc{ZeroShot} costs \$82.13, i.e., about \$0.29--\$0.30 per instance. 
These methods remain cheap because they primarily prompt the model with the issue description and a limited amount of repository context to produce a test, without iterative verification or multi-round refinement. In contrast, agentic methods incur substantially higher costs due to longer contexts analysis and repeated interactions: \textsc{SWE-Agent} costs \$290.71 (about \$1.05 per instance) and \textsc{AutoCodeRover} costs \$368.40 (about \$1.33 per instance). e-Otter++ is a top-tier approach, with per-instance costs of \$1.80 (GPT-4o) and \$2.75 (Claude-3.7-Sonnet) when Agentless-based patch selection is included \cite{ahmed2025execution}. When Agentless is excluded, e-Otter++ costs \$1.10 per instance with GPT-4o and \$1.70 per instance with Claude. AssertFlip~\cite{khatib2025assertflip} reports costs on all 433 verified instances from SWE-bench-V using GPT-4o, with an overall cost of \$435.92, averaging \$1.00 per instance. By comparison, \textsc{Echo} costs \$1.54 per instance using Gemini-2.5 Pro, which is relatively cheaper while achieving higher performance.

\begin{tcolorbox}[title = {Summary of RQ3}]

(1) \textsc{Echo} is relatively cheaper than the competitor while maintaining similar levels of performance.

(2) \textsc{Echo}’s token consumption is dominated by the input tokens. Having a large amount of retrieved context helps generation and analysis, but it is costly.

(3) Solving instances from the \texttt{django} project is the most expensive on average, as it contains several hard issues to reproduce.

\end{tcolorbox}

\section{Discussion}\label{sec:discussion}
In this section we summarize the key findings, pitfalls and lessons learned from \textsc{Echo}.

\textit{Code graph enhanced retrieval yields more useful context.} RQ2 reveals that \textsc{Echo} without refinements still achieves performance higher than Otter++, which uses a similar context to generate the test case but relies on a more naive context retrieval method than \textsc{Echo}. This indicates that using code graphs to guide context retrieval is effective.

\textit{Using execution feedback to refine the generated test case is useful.} The execution feedback not only provides useful information for selecting a test case from the candidate pool, but can also be used to refine test generation. \textsc{Echo} demonstrates that, with this refinement loop, generating just one test case per issue is sufficient to achieve strong performance.

\textit{Automated execution of test cases remains challenging.} We explored using LLMs to automatically execute the generated test cases. Overall, all test cases were eventually executed successfully by our LLM-based test executor, which is promising. However, we observed several pitfalls in practice. Although we defined strict rules to prevent operations that could alter the test environment, the LLM sometimes ignored these constraints and performed actions that changed the environment, causing the test execution to fail.

\textit{How to provide less but sufficiently useful context needs further investigation.} Our cost analysis shows that \textsc{Echo} consumes most of its tokens on long input contexts used for issue analysis. This long input is necessary to ground test generation and subsequent refinement, but it also dominates the overall cost. An important direction for future work is to design context selection strategies, such as learning to predict which focal code, tests, and files are actually needed for a given issue, and dynamically expanding the context only when execution feedback indicates missing dependencies or incorrect assumptions.

\textit{How to measure the correct reported behavior remains challenging.} In this work, we follow e-Otter++ and use generated candidate patches to specify the intended behavior. Although this approach improves issue-reproducing test generation, it remains unreliable for demonstrating the fail-to-pass criterion. The patch may be misleading and does not reflect the error behavior reported in the issue. A robust and correct oracle to demonstrate the failure reported in the issue requires further exploration.

\section{Threats to Validity}\label{sec:threats}
\textit{Internal Validity}:
Our results could be different because we ran the experiments only once. Since LLM outputs can vary with the prompt template and temperature, rerunning the experiments may yield different results. Nevertheless, these results suggest a promising direction for generating issue-reproducing tests.
\noindent \textit{External Validity}: LLMs are evolving rapidly, and we did not evaluate \textsc{Echo} with the latest models (e.g., Gemini 3.0 Pro). Given that newer and more capable models typically achieve stronger results, we expect \textsc{Echo} to perform better with a more advanced model.

\section{Conclusion and Future Work}\label{sec:conclusion}
In this paper, we proposed \textsc{Echo}, a system for generating issue-reproducing test cases based on issues.
In comparison to existing approaches for issue reproduction test generation, \textsc{Echo} is novel in that it (1) enhances the information retrieval phases by using a code graph and LLM-based auto prompt refinement to retrieve sufficient context, (2) automatically runs test cases, (3) generates candidate patches, using the patched version of the project to verify the fail-to-pass criterion and refine the test based on execution feedback if necessary, and (4) generates just one high quality test case per instance, thus ensuring efficiency. 
Through extensive experiments, we demonstrated that \textsc{Echo}'s superior performance can primarily be attributed to the retrieval of sufficient and useful context %
as well as the verification and iterative refinement mechanisms based on the potential patch version. These novel ideas collectively enabled \textsc{Echo} to outperform previous work and become the new SOTA on the widely-used SWT-Bench Verified dataset. In future work, we plan to explore (1) how to reduce the retrieval context to improve efficiency, (2) find better ways to run the generated test cases, and (3) explore new oracles to better demonstrate the failures reported in the issues.

\section*{Data Availability}
Our code, results, and all the execution logs are available at \url{https://github.com/EuniAI/Echo}.

\bibliographystyle{ACM-Reference-Format}
\bibliography{sample-base}

\appendix

\end{document}